\documentstyle[12pt,epsf]{article}
\begin{document}
\begin{titlepage}
\title{Thomas-Ehrman shifts in nuclei around $^{16}$O
and role of residual nuclear interaction}
\author{K. Ogawa\footnote{E-mail: ogawa@c.chiba-u.ac.jp},
H. Nakada\footnote{E-mail: nakada@c.chiba-u.ac.jp},
S. Hino and R. Motegi\\
{\it Department of Physics, Chiba University,
Chiba 263-8522, Japan}}
\date{\today}
\maketitle
\thispagestyle{empty}
\begin{abstract}
The asymmetry in the energy spectra between mirror nuclei
(the Thomas-Ehrman shifts) around $^{16}$O
is investigated from a phenomenological viewpoint.
The recent data on proton-rich nuclei indicates
that the residual nuclear interaction is reduced
for the loosely bound $s$-orbit by as much as 30\%,
which originates in the broad radial distribution
of the proton single-particle wave function.
\end{abstract}
\noindent
PACS numbers: 21.10.-k, 27.20.+n, 21.60.Cs
\vspace*{3mm}\noindent
Keywords: Thomas-Ehrman shift, mirror nuclei, proton-rich nuclei,
residual interaction, loosely bound $s$-orbit, shell model
\end{titlepage}
\pagestyle{plain}
\section{Introduction}
Structures of  proton-rich nuclei are important
for the rapid-proton ($rp$) process of the nucleosynthesis,
which takes place in the hydrogen burning stage in stellar site.
Since the strong interaction keeps the charge symmetry very well
and the Coulomb energies are almost state-independent in a nuclide,
energy spectra are quite analogous between mirror nuclei.
Hence we usually estimate the level structures of $Z>N$ exotic nuclei
from their mirror partners. 
However, for example, the excitation energies of the $1/2^+$ first
excited states in $^{13}$C and $^{13}$N show large discrepancy,
which is called Thomas-Ehrman shift (TES)\cite{ref:TE}\footnote{
There has been a confusion in the term `Thomas-Ehrman shift'.
In some references it was used in a restrictive meaning
of a specific effect of the Coulomb force\cite{ref:Au83}.
In this paper we use it in more general sense of
the level shift between mirror nuclei.}.
The TES may have a significant influence
on the scenario of the $rp$ process,
and it is highly desired to predict the TES correctly.
Recent experiments provide us with valuable information
of energy levels of $Z>N$ nuclei around $^{16}$O.
The TES has conventionally been regarded
as an effect of the Coulomb force
on a loosely bound or unbound proton occupying an $s$-orbit.
With the aid of the recent data,
it is being possible to argue
whether this mechanism is sufficient to account for the TES
in various mirror nuclei.
The difference in the single-particle (s.p.) energies
leads to a certain difference in the s.p. wave functions
between protons and neutrons, in general.
This affects the matrix elements of residual nuclear interaction
(RNI), even though the original $NN$ force is charge symmetric.
The question is whether this effect on the TES is sizable.
Since the nuclear interaction has short-range character,
it is expected that the RNI becomes weaker
as the s.p. wave functions distributes over a wider region.
The RNI reduction due to the broad radial distribution
of the s.p. wave functions typically amounts
only to a few percent\cite{ref:CSB},
which does not cause notable TES for low-lying states.
However, because a loosely bound $s$-orbit can have very long tail
owing to the lack of the centrifugal barrier,
the levels involving such an $s$-orbit may have
substantial contribution of the RNI to the TES.
In this paper we investigate the TES in $A\sim 16$ nuclei
from a phenomenological viewpoint,
primarily focusing on effects of the RNI on the TES,
via the data of the $^{16}$N--$^{16}$F, $^{15}$C--$^{15}$F and
$^{16}$C--$^{16}$Ne mirror pairs.
\section{Phenomenological study of Thomas-Ehrman shifts
around $^{16}$O}
\subsection{$^{16}$N--$^{16}$F}
We shall take the $(0p_{1/2})^{-n_1}\otimes 
(0d_{5/2} 1s_{1/2})^{n_2}$
model space on top of the $^{16}$O inert core.
For neutron-rich nuclei with $Z\le 8\le N$,
$n_1=8-Z$ and $n_2=N-8$,
and vice versa for their mirror partners.
The single-particle (hole) energies are obtained
from the data of $^{17}$O and $^{17}$F 
($^{15}$N and $^{15}$O)\cite{ref:TI}.
Taking into account their mass differences 
from $^{16}$O\cite{ref:mass}, we have (in MeV)
\begin{eqnarray} 
\epsilon_{\rm n}(0d_{5/2})=-4.144, \ \ \ 
\epsilon_{\rm n}(1s_{1/2})=-3.273,\nonumber\\
\epsilon_{\rm p}(0d_{5/2})=-0.600, \ \ \  
\epsilon_{\rm p}(1s_{1/2})=-0.105, \label{eq:spe-p}
\end{eqnarray}
and
\begin{equation} 
\epsilon_{\rm p}(0p_{1/2}^{-1})=12.128, \ \ \ 
\epsilon_{\rm n}(0p_{1/2}^{-1})=15.664. \label{eq:spe-h}
\end{equation}
The shift in $E_x(1/2^+)$ of $^{17}$F
(i.e. $\Delta\epsilon_{\rm p}^{s-d}\equiv
\epsilon_{\rm p}(1s_{1/2})-\epsilon_{\rm p}(0d_{5/2})
=0.495$~MeV) from that of $^{17}$O
(i.e. $\Delta\epsilon_{\rm n}^{s-d}\equiv
\epsilon_{\rm n}(1s_{1/2})-\epsilon_{\rm n}(0d_{5/2})
=0.871$~MeV) is a typical TES.
Because the proton in the $1s_{1/2}$ orbit is loosely bound
and free from the influence of the centrifugal barrier,
its wave function spreads in a radial direction
(like halo or skin structure),
leading to weaker Coulomb repulsion than that of $0d_{5/2}$.
This difference in the Coulomb energy gives rise to the TES
for such a core plus one-particle system\cite{ref:TE}.
The mechanism how the energy shift of $\Delta\epsilon_{\rm p}^{s-d}$
from $\Delta\epsilon_{\rm n}^{s-d}$ occurs
has recently been investigated in some detail in Ref.\cite{ref:AKI}.
The observed energy spectra of $^{16}$N and $^{16}$F,
the $T_z=\pm 1$ mirror nuclei,
show a remarkable difference\cite{ref:TI}.
Even the ground state spins do not match,
being $2^{-}$ in $^{16}$N and $0^{-}$ in $^{16}$F. 
On top of the $^{16}$O core, the lowest four states 
in these nuclei ($2^-, 0^-, 3^-, 1^-$ in $^{16}$N and $0^-, 1^-, 
2^-, 3^-$ in $^{16}$F) are classified
into the $|0p_{1/2}^{-1}0d_{5/2};J=2^{-},3^{-} \rangle$ and 
$|0p_{1/2}^{-1}1s_{1/2};J=0^{-},1^{-} \rangle$ multiplets.
The difference in low-lying energy spectra
is linked to the relatively low energy
of the proton $1s_{1/2}$ orbit.
The smaller $\Delta\epsilon_{\rm p}^{s-d}$
than $\Delta\epsilon_{\rm n}^{s-d}$ in Eq.~(\ref{eq:spe-p})
shifts down the $0^-$ and $1^-$ states of $^{16}$F.
However, the amount of the observed TES in the $0^-$ and $1^-$ states
is $\simeq 0.70$~MeV on average,
notably larger than $\Delta\epsilon_{\rm n}^{s-d}
- \Delta\epsilon_{\rm p}^{s-d} = 0.376$~MeV.
It is likely that the two-body RNI
has a substantial contribution to this TES.
Since the last proton is unbound in $^{16}$F while bound in $^{17}$F,
the relative energy of $(1s_{1/2})_{\rm p}$ may be further lowered
in $^{16}$F,
mainly by the effect of the Coulomb force.
However, it is not simple to evaluate separately the RNI
and the nucleus-dependence of $\epsilon_{\rm p}(1s_{1/2})$
(or $\Delta\epsilon_{\rm p}^{s-d}$).
We here assume the $\epsilon$'s 
of Eqs.~(\ref{eq:spe-p},\ref{eq:spe-h}),
ignoring the nucleus-dependence of $\Delta\epsilon_{\rm p}^{s-d}$,
whose argument will be given later.
Then the matrix elements of the residual proton-neutron interaction
between a $0p_{1/2}$ hole and an $s,d$ particle can be derived
from the experimental levels of $^{16}$N and $^{16}$F.
The results are presented in Table~\ref{tab:Vpn}.
We here denote the diagonal matrix element
$\langle (j_1)_\rho (j_2)_\tau; J | V
 | (j_1)_\rho (j_2)_\tau; J \rangle$
by $V_{\rho\tau}(j_1 j_2;J)$,
where $(\rho,\tau)=({\rm p},{\rm n})$.
For example, $V_{\rm pn}(0p_{1/2}^{-1}0d_{5/2};J)=
\langle (0p_{1/2}^{-1})_{\rm p} (0d_{5/2})_{\rm n}; J | V
| (0p_{1/2}^{-1})_{\rm p} (0d_{5/2})_{\rm n}; J \rangle$
and $V_{\rm np}(0p_{1/2}^{-1}0d_{5/2};J)= 
\langle (0p_{1/2}^{-1})_{\rm n} (0d_{5/2})_{\rm p}; J | V
| (0p_{1/2}^{-1})_{\rm n} (0d_{5/2})_{\rm p}; J \rangle$.
While the matrix elements 
$V_{\rm np}(0p_{1/2}^{-1}0d_{5/2};J)$
deduced from $^{16}$F are similar to
$V_{\rm pn}(0p_{1/2}^{-1}0d_{5/2};J)$ from $^{16}$N,
the elements regarding $1s_{1/2}$,
$V_{\rm pn}(0p_{1/2}^{-1}1s_{1/2};J)$ and
$V_{\rm np}(0p_{1/2}^{-1}1s_{1/2};J)$,
show obvious discrepancy.
The $V_{\rm np}(0p_{1/2}^{-1}1s_{1/2};J)$ element
is smaller by a factor of about 0.7 than
$V_{\rm pn}(0p_{1/2}^{-1}1s_{1/2};J)$,
both for $J=0^-$ and $1^-$.
The reduction of the proton-neutron RNI matrix elements
indicates that the TES in the $^{16}$N--$^{16}$F pair
owes a part to the nuclear force,
not only to the Coulomb force which relatively shifts down
$\epsilon_{\rm p}(1s_{1/2})$.
%Moreover, the charge symmetry breaking in the free $NN$ interaction
%is unable to lead to the difference
%between $V_{\rm np}$ and $V_{\rm pn}$.
The reduction of $V_{\rm np}$ compared with $V_{\rm pn}$
should originate in the difference
of the s.p. radial wave functions
between a proton and a neutron.
The $V_{\rm pp}$--$V_{\rm nn}$ difference
in the RNI has been investigated in a wide mass region\cite{ref:CSB},
which yields a few percent reduction of $V_{\rm pp}$
relative to $V_{\rm nn}$,
as a result of the proton--neutron difference
in the s.p. wave functions.
This coincides with the $V_{\rm np}/V_{\rm pn}$ ratio
for $0d_{5/2}$ shown in Table~\ref{tab:Vpn}.
On the other hand, the present RNI reduction
with respect to $1s_{1/2}$ is remarkably stronger
than the global systematics.
The strong reduction of the RNI is possibly an effect
of the loosely bound $s$-orbit.
Because of the lack of the centrifugal barrier,
the radial function of the $s$-orbit
depends appreciably on the separation energy.
Since the proton $1s_{1/2}$ orbit is loosely bound
in the proton-rich nuclei of this mass region,
the $1s_{1/2}$ proton wave function $R_{1s_{1/2}}(r_{\rm p})$
distributes with a long tail and
its spatial overlap with another nucleon is depressed,
in comparison with $R_{1s_{1/2}}(r_{\rm n})$.
Therefore nuclear force is expected to give
appreciably smaller matrix elements
if loosely bound or unbound protons are involved.
We shall examine this mechanism in Section~\ref{sec:WS+M3Y}.
It is noted that the residual interaction
relevant to the low-lying levels is repulsive
in $^{16}$N--$^{16}$F, because of the particle-hole conjugation.
In addition to the small $\Delta\epsilon_{\rm p}^{s-d}$
relative to $\Delta\epsilon_{\rm n}^{s-d}$,
the reduction of the repulsive RNI lowers
the levels involving $(1s_{1/2})_{\rm p}$,
relative to those having $(0d_{5/2})_{\rm p}$.
Thus the TES is enhanced in this pair of mirror nuclei.
We shall investigate other mirror pairs
based on the above phenomenological Hamiltonian.
The validity of the current approach would be
tested via the systematic study of this sort,
whereas for the time being we put aside the possibility
of the nucleus-dependence of $\epsilon$'s.
As will be shown, the systematic study supports
the RNI reduction picture on the TES.
\subsection{$^{15}$C--$^{15}$F}
\begin{figure}
\epsfysize=9.0cm
\centerline{\epsffile{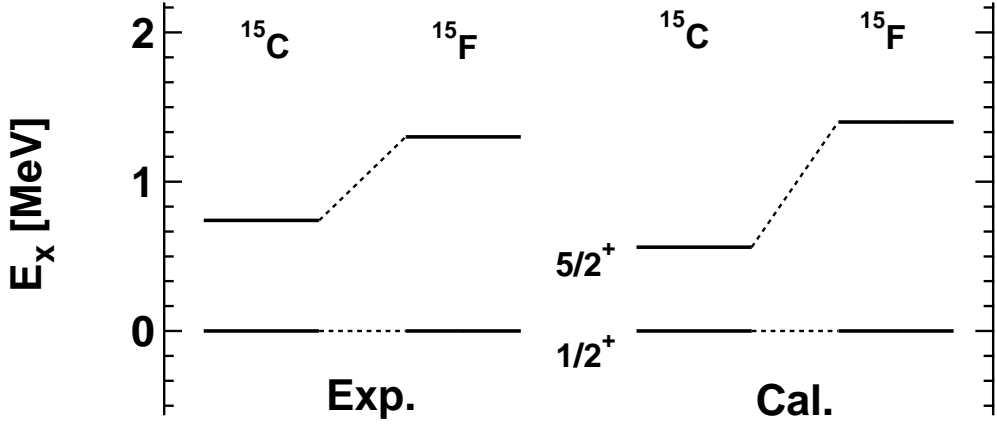}}
\vspace{2mm}
\caption{Experimental and calculated energy spectra of
the $^{15}$C--$^{15}$F mirror nuclei.}
\label{fig:15C-F}
\end{figure}

By using the empirical $\epsilon$'s of
Eqs.~(\ref{eq:spe-p},\ref{eq:spe-h})
and the $\langle V\rangle$'s of Table~\ref{tab:Vpn},
we calculate energies of the low-lying levels $5/2^+$ and $1/2^+$
of $^{15}$C and $^{15}$F ($T_z=\pm 3/2$)
in the $(0p_{1/2})^{-2}\otimes(0d_{5/2}1s_{1/2})^1$ model space.
The calculated energy levels are shown in Fig.~\ref{fig:15C-F}
together with the experimental data\cite{ref:TI}.
The level inversion occurs;
the $1/2^+$ states, instead of $5/2^+$,
become lowest for both nuclei.
This inversion is reproduced by the shell model Hamiltonian,
due to the repulsion shown in Table~\ref{tab:Vpn}
which is stronger between $0p_{1/2}^{-1}$ and $0d_{5/2}$
than between $0p_{1/2}^{-1}$ and $1s_{1/2}$.
The $5/2^+$ level is observed at $E_x=0.740$~MeV in $^{15}$C,
while at 1.300~MeV in $^{15}$F.
The shell model yields $E_x(5/2^+)=0.563$~MeV
for $^{15}$C and 1.400~MeV for $^{15}$F.
The TES in the $^{15}$C--$^{15}$F pair is thus described
with a reasonable accuracy, though slightly overshot,
within the framework of the phenomenological shell model.
Weaker repulsion in $V_{\rm np}(0p_{1/2}^{-1}1s_{1/2};J)$
than in $V_{\rm pn}(0p_{1/2}^{-1}1s_{1/2};J)$
plays an appreciable role in the TES.
The $V_{\rm pp}(0p_{1/2}^{-2};J=0^+)$ and
$V_{\rm nn}(0p_{1/2}^{-2};J=0^+)$ matrix elements,
which do not affect the excitation energies,
can be evaluated from the ground-state energies
of $^{14}$C and $^{14}$O.
We can then calculate the absolute values of the energies,
not only the excitation energies,
of the $^{15}$C and $^{15}$F levels.
The biggest discrepancy is found in $E(1/2^+)$ of $^{15}$C,
which is overestimated by 0.166~MeV,
whereas the other energies are reproduced 
within the 0.1~MeV accuracy.
This may suggest that an additional effect is missed
for the $1s_{1/2}$ neutron,
whose separation energy is small (1.218~MeV) in $^{15}$C.
\subsection{$^{16}$C--$^{16}$Ne}
\begin{figure}
\epsfysize=12.0cm
\centerline{\epsffile{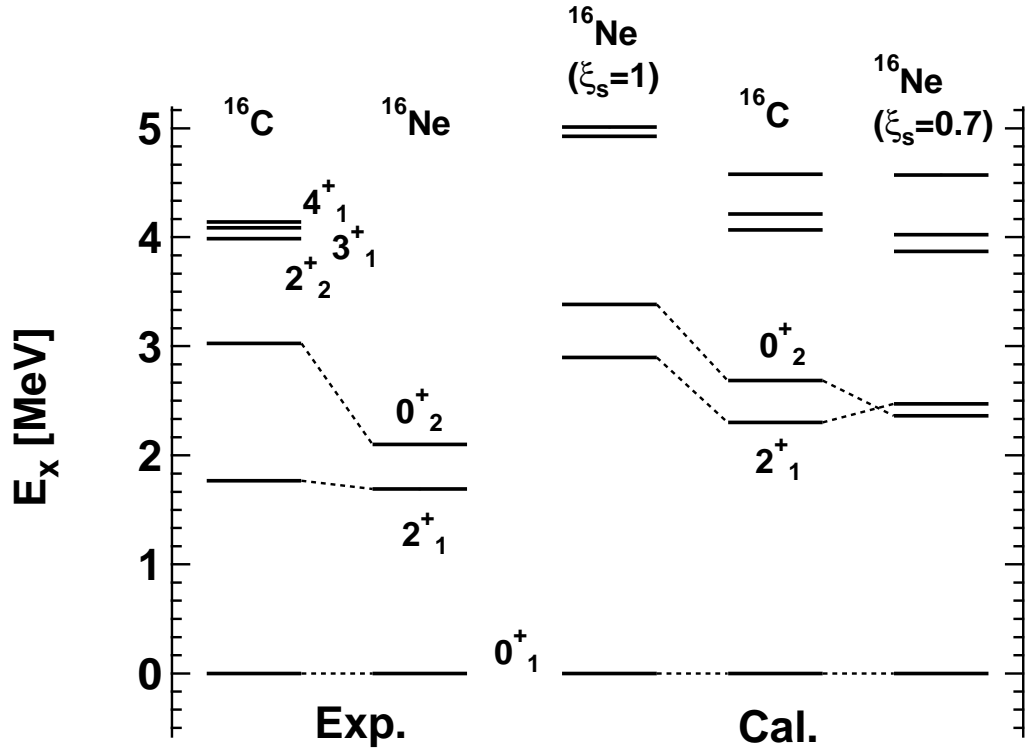}}
\vspace{2mm}
\caption{Experimental and calculated (with and without
the reduction factor $\xi_s$) energy spectra of
the $^{16}$C--$^{16}$Ne mirror nuclei.}
\label{fig:16C-Ne}
\end{figure}

The $^{16}$C--$^{16}$Ne pair ($T_z=\pm 2$) is significant as well,
in investigating the effect of the RNI on the TES.
The low-lying states of these nuclei have the $0p_{1/2}^{-2}
\otimes(0d_{5/2}1s_{1/2})^2$ configuration.
Since the $0p_{1/2}^{-2}$ part
does not contribute to the excitation energy,
the TES can disclose difference between proton-proton ($V_{\rm pp}$)
and neutron-neutron ($V_{\rm nn}$) interactions in the $sd$-shell.
As an effective interaction in the $sd$-shell,
the so-called USD interaction\cite{ref:USD}
has successfully been used.
Although the USD interaction is derived for the full $sd$-shell
calculation,
we neglect the $0d_{3/2}$ components,
since the $0d_{3/2}$ orbit is hardly occupied in low-lying states
of the nuclei around $^{16}$O.
Indeed, we can well reproduce the low-lying levels of $^{18}$O
with the USD interaction in the $(0d_{5/2}1s_{1/2})^2$ space.
We carry out the shell model calculation
in the $0p_{1/2}^{-2}\otimes(0d_{5/2}1s_{1/2})^2$ space,
with the Hamiltonian comprised of the empirical $\epsilon$'s
and $\langle V\rangle$'s
(see Eqs.~(\ref{eq:spe-p},\ref{eq:spe-h}) and Table~\ref{tab:Vpn})
as well as of the USD matrix elements.
The binding energy of $^{16}$C is reproduced
with the accuracy of about 0.1~MeV.
In computing the binding energy of $^{16}$Ne,
the residual two-body Coulomb interaction has to be estimated.
With the s.p. wave functions in the Woods-Saxon potential
which will be mentioned in Section~\ref{sec:WS+M3Y},
the two-body Coulomb force yields approximately constant energy shift
of about 0.4~MeV for low-lying levels,
within the accuracy of 0.1~MeV.
If we use the charge-symmetric (i.e. $V_{\rm pp}=V_{\rm nn}$)
USD interaction with this Coulomb correction,
the binding energy of $^{16}$Ne is seriously overestimated
by as much as 0.8~MeV.
Because of the level inversion in $^{15}$C and $^{15}$F,
$1s_{1/2}$ lies lower than $0d_{5/2}$ in an effective sense.
Thereby the ground state consists mainly of
the $0p_{1/2}^{-2}\otimes 1s_{1/2}^2$ configuration,
with small admixture of $0p_{1/2}^{-2}\otimes 0d_{5/2}^2$.
It is likely for the RNI matrix elements
involving $(1s_{1/2})_{\rm p}$
to suffer some amount of reduction,
because the $1s_{1/2}$ protons are bound loosely (or unbound).
For this reason we reduce the USD matrix elements
concerning the $(1s_{1/2})_{\rm p}$ orbit 
by an overall factor $\xi_s$,
while not changing the other matrix elements.
The binding energy of $^{16}$Ne is found to be reproduced
if we set $\xi_s\simeq 0.7$ (i.e.
$V_{\rm pp}(0d_{5/2}1s_{1/2};J)\simeq
0.7\times V_{\rm nn}(0d_{5/2}1s_{1/2};J)$, and so forth).
It is notable that this factor is close to the proton-neutron ratio
$V_{\rm np}/V_{\rm pn}$ concerning $1s_{1/2}$ in Table~\ref{tab:Vpn}.
Recently $E_x(0^+_2)$ of $^{16}$Ne has been reported\cite{ref:Ne16},
indicating a large TES.
$E_x(0^+_2)$ of $^{16}$Ne is lower by about 1~MeV 
than that of $^{16}$C.
In Fig.~\ref{fig:16C-Ne} the results of the $\xi_s=1$
(i.e. no reduction of the RNI) and $\xi_s=0.7$
cases are compared with the experimental data.
As has been noticed,
the $1s_{1/2}$ energy relative to $\epsilon(0d_{5/2})$
is deeper for protons than for neutrons.
This tends to lower the ground state of $^{16}$Ne,
whose main configuration is $0p_{1/2}^{-2}\otimes 1s_{1/2}^2$.
Thus, if we use the charge-symmetric USD interaction
(i.e. $\xi_s=1$),
$E_x(0^+_2)$ of $^{16}$Ne becomes necessarily higher
than that of $^{16}$C.
The recent data of $E_x(0^+_2)$ clearly favors the reduction
of the RNI regarding the $(1s_{1/2})_{\rm p}$ orbit.
The reduction with $\xi_s=0.7$ almost reproduces
$E_x(0^+_2)$ of $^{16}$Ne, as well as the binding energy.
Note that the residual Coulomb force does not contribute
seriously to the TES, as far as the number of the valence protons
is not large.
With this reduction of $\xi_s=0.7$ the known energy spectra
of the $^{17}$Ne\cite{ref:Ne17} are also reproduced.
\subsection{Discussion --- 
$\Delta\epsilon_{\rm p}^{s-d}$ decrease vs. RNI reduction}
We now consider the possibility
of the nucleus-dependence of $\Delta\epsilon_{\rm p}^{s-d}
= \epsilon_{\rm p}(1s_{1/2})-\epsilon_{\rm p}(0d_{5/2})$.
In extracting the RNI matrix elements from $^{16}$N--$^{16}$F,
we have assumed the s.p. energies taken
from the $^{17}$O--$^{17}$F data.
One may argue that in the $^{16}$F nucleus
$(1s_{1/2})_{\rm p}$ receives less Coulomb repulsion
than in $^{17}$F, because the last proton is unbound,
and that the TES in $^{16}$N--$^{16}$F should be ascribed to
the corresponding lowering of $\epsilon_{\rm p}(1s_{1/2})$
relative to $\epsilon_{\rm p}(0d_{5/2})$,
instead of the reduction of the RNI.
As far as we view only the $^{16}$N--$^{16}$F data,
the RNI reduction does not seem to be an exclusive solution.
In this regard, the TES in the $^{16}$C--$^{16}$Ne pair
has particular importance.
Since $^{16}$Ne has negative $S_{2p}$ (two proton separation energy),
$\epsilon_{\rm p}(1s_{1/2})$ 
(relative to $\epsilon_{\rm p}(0d_{5/2})$)
goes down in $^{16}$Ne from the value obtained in $^{17}$F,
if we follow the argument of the nucleus-dependence
of $\Delta\epsilon_{\rm p}^{s-d}$ as in $^{16}$F.
However, this makes $E_x(0^+_2)$ in $^{16}$Ne even higher
than in the $\xi_s=1$ case of Fig.~\ref{fig:16C-Ne},
where we already have the wrong sign of the TES.
It is noted that the RNI derived from $^{16}$N--$^{16}$F
(shown in Table~\ref{tab:Vpn}) is repulsive
because of the particle-hole nature,
while the $sd$-shell interaction crucial to the TES
in the $^{16}$C--$^{16}$Ne pair is attractive.
Thereby, the $\Delta\epsilon_{\rm p}^{s-d}$ decrease
and the RNI reduction gives opposite effects
on the loosely bound (or unbound) $s$-orbit in $^{16}$C--$^{16}$Ne.
If only the $\Delta\epsilon_{\rm p}^{s-d}$ decrease is considered,
obvious contradiction to the data results.
Thus, lower $E_x(0^+_2)$ in $^{16}$Ne
than in $^{16}$C cannot be described
without the reduction of the RNI for $(1s_{1/2})_{\rm p}$.
The data implies that, although the nucleus-dependence
of $\Delta\epsilon_{\rm p}^{s-d}$ may exist,
its effect on the TES seems much less significant
than that of the RNI reduction.
On the contrary, the reduction of the RNI naturally accounts for
the TES's around $^{16}$O, in particular those of $^{16}$N--$^{16}$F
and $^{16}$C--$^{16}$Ne, simultaneously.
\subsection{$E_x(3^+)$ of $^{18}$Ne}
The TES is not apparent for the lowest-lying states
of the $^{18}$O--$^{18}$Ne mirror pair,
since their main configuration is $0d_{5/2}^2$.
On the other hand, the lowest $3^+$ state,
which is observed at $E_x=5.378$~MeV in $^{18}$O,
is expected to have the $0d_{5/2}1s_{1/2}$ configuration.
Because the $1s_{1/2}$ orbit is relevant,
the TES for this state may be sizable.
The energy of the $3^+$ state of $^{18}$Ne is important
in the scenario of the $rp$ process\cite{ref:WGT}.
Whereas an earlier experiment gives $E_x=4.56$~MeV\cite{ref:Ne18_3+},
this $3^+$ state has not been established
by recent experiments\cite{ref:Ne18}.
By using the USD interaction
with the reduction factor $\xi_s=0.7$ for $(1s_{1/2})_{\rm p}$
(together with $\epsilon_{\rm p}$'s deduced from $^{17}$F),
we evaluate TES of 0.86~MeV for this $3^+$ state.
This leads to $E_x(3^+_1)\simeq 4.5$~MeV,
in good agreement with Ref.\cite{ref:Ne18_3+}.
\section{Mechanism of RNI reduction}
\label{sec:WS+M3Y}
\begin{figure}
\epsfxsize=16.0cm
%\epsfysize=12.0cm
%\centerline{\epsffile{Vrat.eps}}
\epsffile{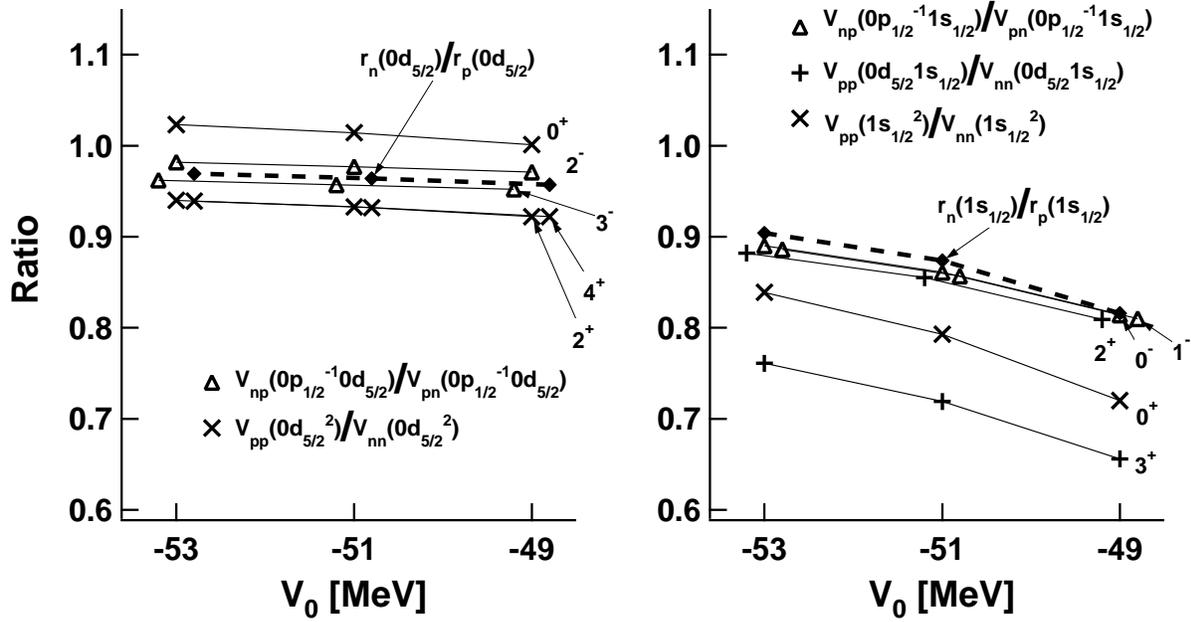}
\vspace{2mm}
\caption{Proton-neutron ratios of the RNI diagonal matrix elements
in the WS+M3Y model, for the WS potential depth varied
around $V_0=-51$~MeV.
The ratios involving the $1s_{1/2}$ orbit
are shown at the right panel,
and those without $1s_{1/2}$ but with $0d_{5/2}$
are at the left panel.
The $J$ values of the two-body states are indicated in the graph.
The corresponding ratios with different $V_0$
are connected by thin lines.
The proton-neutron ratios of the rms radii of the s.p. orbits
are also shown for $j=0d_{5/2}$ (left panel)
and $1s_{1/2}$ (right panel),
by filled diamonds linked by thick dashed lines.
}
\label{fig:WS-M3Y}
\end{figure}

We next study the mechanism of the RNI reduction
concerning the proton $1s_{1/2}$ orbit,
from a qualitative (or semi-quantitative) standpoint.
As has been pointed out, it is likely
that the RNI reduction is connected
to the broad distribution of $R_{1s_{1/2}}(r_{\rm p})$.
The amount of the reduction, however, is notably large,
compared with the global trend
which has been estimated to be a few percent\cite{ref:CSB}.
It is a question whether $R_{1s_{1/2}}(r_{\rm p})$
distributes so broadly as to give RNI reduction by about 30\%,
despite the presence of the Coulomb barrier.
It is not an easy task to estimate microscopically
the RNI matrix elements to a good precision.
Instead, we view proton-neutron ratio of the RNI matrix elements
($V_{\rm np}/V_{\rm pn}$ and $V_{\rm pp}/V_{\rm nn}$)
of the M3Y interaction.
The M3Y force\cite{ref:M3Y} basically represents the $G$-matrix
and therefore somewhat realistic,
and enables us to avoid tedious computation.
To take into account effects of the loose binding
with the centrifugal barrier and the Coulomb barrier,
the single-particle wave functions are obtained
under the Woods-Saxon (WS) plus Coulomb potential.
We adopt the WS parameters of Ref.~\cite{ref:WS} at $^{16}$O,
varying the WS potential depth $V_0$
around the normal value $-51~{\rm MeV}$.
Even if the absolute values of the RNI matrix elements
are not reliable,
the proton-neutron ratios carries certain information,
because they depend mainly on the proton-neutron difference
of the s.p. wave functions.
Note that core polarization effects are not taken into consideration
in this WS+M3Y calculation,
which should be included in the shell model interaction.
The present proton-neutron ratios give
only qualitative (or semi-quantitative) nature of the RNI,
since they do not need to match precisely the shell model ones
(for instance, Table~\ref{tab:Vpn}).
The WS potential with $V_0=-53$ gives (in MeV)
\begin{eqnarray}
\epsilon_{\rm n}(0d_{5/2})=-7.52, \ \ \ 
\epsilon_{\rm n}(1s_{1/2})=-4.80,\nonumber\\
\epsilon_{\rm p}(0d_{5/2})=-3.90, \ \ \  
\epsilon_{\rm p}(1s_{1/2})=-1.46. \label{eq:spe-v53}
\end{eqnarray}
As the potential becomes shallower,
the s.p. orbits are bound more loosely.
Indeed, at $V_0=-51$ we have
\begin{eqnarray}
\epsilon_{\rm n}(0d_{5/2})=-6.36, \ \ \ 
\epsilon_{\rm n}(1s_{1/2})=-3.98,\nonumber\\
\epsilon_{\rm p}(0d_{5/2})=-2.80, \ \ \  
\epsilon_{\rm p}(1s_{1/2})=-0.76, \label{eq:spe-v51}
\end{eqnarray}
and at $V_0=-49$
\begin{eqnarray}
\epsilon_{\rm n}(0d_{5/2})=-5.23, \ \ \ 
\epsilon_{\rm n}(1s_{1/2})=-3.22,\nonumber\\
\epsilon_{\rm p}(0d_{5/2})=-1.75, \ \ \  
\epsilon_{\rm p}(1s_{1/2})=-0.14. \label{eq:spe-v49}
\end{eqnarray}
Whereas the wave function is insensitive to $\epsilon$
for the deeply bound orbits,
it is not the case for the loosely bound orbit $(1s_{1/2})_{\rm p}$.
The variation of the wave function is typically measured
by the mean radius of the orbit
$r_\rho(j)\equiv \sqrt{\langle (j)_\rho|r^2|(j)_\rho\rangle}$
($\rho={\rm p}, {\rm n}$).
By varying the WS parameter $V_0$,
we see how the RNI as well as $r_\rho(j)$ behave 
as $\epsilon$ changes.
For the M3Y matrix elements,
the proton-neutron ratios $V_{\rm np}/V_{\rm pn}$
with respect to the $(0p_{1/2})^{-1}\otimes(0d_{5/2}1s_{1/2})^1$
two-body states
and $V_{\rm pp}/V_{\rm nn}$ with respect to the
$(0d_{5/2}1s_{1/2})^2$ states
are depicted in Fig.~\ref{fig:WS-M3Y}.
Though the ratios of the off-diagonal elements
$\langle (1s_{1/2}^2)_\rho;0^+|V|(0d_{5/2}^2)_\rho;0^+ \rangle$
and $\langle (0d_{5/2}1s_{1/2})_\rho;2^+|V
|(0d_{5/2}^2)_\rho;2^+ \rangle$ are not shown,
they behave in a similar manner
to those of diagonal elements involving $1s_{1/2}$.
The proton-neutron ratios of the rms radii
of the s.p. orbits are also presented
in Fig.~\ref{fig:WS-M3Y} for $j=0d_{5/2}$ and $1s_{1/2}$.
As is expected,
$R_{1s_{1/2}}(r_{\rm p})$ distributes over a broader region
than $R_{1s_{1/2}}(r_{\rm n})$, to a certain extent.
In Fig.~\ref{fig:WS-M3Y}, the rms radius of $(1s_{1/2})_{\rm p}$
is larger by about 10--20\% than that of $(1s_{1/2})_{\rm n}$,
somewhat depending on $V_0$.
In contrast, the rms radius of $(0d_{5/2})_{\rm p}$ differ
only by a few percent from that of $(0d_{5/2})_{\rm n}$,
insensitive to $V_0$.
From Fig.~\ref{fig:WS-M3Y} we confirm the following two points:
(a) the RNI reduction well correlates
to the increase of the rms radii of the relevant orbits,
and (b) the matrix elements involving $(1s_{1/2})_{\rm p}$
can be reduced from those of $(1s_{1/2})_{\rm n}$
by as much as a few tens percent around $^{16}$O.
The former point is consistent with Ref.\cite{ref:Kuo},
though we use more realistic s.p. wave functions
(but less realistic $G$-matrix) than in Ref.\cite{ref:Kuo}.
The latter implies that the broad distribution
of $R_{1s_{1/2}}(r_{\rm p})$ seems accountable
for the RNI reduction.
Although there remain additional interests
in the RNI reduction (e.g. accurate estimate of the reduction factor,
nucleus- and/or state-dependence of the reduction factor),
they will require reliable treatment 
of the core polarization effects,
which is beyond the scope of the current study.
We just point out at this moment that,
due to the broad distribution of the s.p. wave function,
the core polarization effects tend to diminish\cite{ref:Kuo}
and therefore the shell model interaction may be reduced further.
The residual Coulomb force hardly contributes
to the excitation energies of low-lying states,
for the nuclei around $^{16}$O,
as far as the number of valence protons remains small.
The state-dependence of the residual Coulomb force
is less than 0.1~MeV for the nuclei under discussion,
if estimated with the above WS wave functions.
\section{Summary}
The Thomas-Ehrman shifts generally occur in the $A\sim 16$ region,
where the $1s_{1/2}$ proton is unbound or loosely bound.
As well as the difference between 
$\Delta\epsilon_{\rm n}^{s-d}$ and 
$\Delta\epsilon_{\rm p}^{s-d}$,
the reduction of the residual nuclear interaction matrix elements
involving the $1s_{1/2}$ proton plays an important role in the TES.
As has been deduced from the nuclei $^{16}$N and $^{16}$F,
the matrix elements
$V_{\rm np}(0p_{1/2}^{-1} 1s_{1/2})$ is notably smaller
than $V_{\rm pn}(0p_{1/2}^{-1} 1s_{1/2})$,
by a factor of about 0.7.
This factor is remarkably smaller than the general trend
of the proton-neutron asymmetry in the RNI.
Similar reduction of $V_{\rm pp}$ in the $sd$-shell
(relative to $V_{\rm nn}$)
accounts for the TES in $E_x(0^+_2)$ of the $^{16}$C--$^{16}$Ne pair
as well as the mass of $^{16}$Ne.
It is remarked that the RNI reduction is far more significant
than the nucleus-dependence of $\Delta\epsilon_{\rm p}^{s-d}$,
as is argued in connection to $E_x(0^+_2)$ of $^{16}$Ne.
Taking into account the RNI reduction,
the TES's observed in $^{15}$C--$^{15}$F and other pairs are
understood within the phenomenological shell model.
On the same ground the astrophysically important $E_x(3^+_1)$
of $^{18}$Ne is predicted to be $\sim 4.5$~MeV.
The reduction of the residual interaction seems to originate in
the broad radial distribution of wave function 
of the $1s_{1/2}$ proton,
which is bound loosely (or unbound)
and is not affected by the centrifugal barrier.
This picture is supported by viewing the proton-neutron ratio
of the M3Y interaction matrix elements,
which are evaluated with the single-particle wave functions
under the Woods-Saxon plus Coulomb potential.
\\
\noindent
Discussions with S. Kubono, K. Kat\={o} and S. Aoyama
are gratefully acknowledged.

\clearpage
\begin{table}
\begin{center}
\caption{Matrix elements of residual proton-neutron interaction
$V_{\rm pn}(j_1 j_2; J)$ and $V_{\rm np}(j_1 j_2; J)$
deduced from $^{16}{\rm N}$ and $^{16}{\rm F}$ (MeV),
and their ratio.\label{tab:Vpn}}
\begin{tabular}{cccrrr}
\hline\hline
$j_1$ & $j_2$ & $J^P$ & $V_{\rm pn}(j_1 j_2; J)$
& $V_{\rm np}(j_1 j_2; J)$ & $V_{\rm np}/V_{\rm pn}$ \\ \hline
$0p_{1/2}^{-1}$ & $0d_{5/2}$ & $2^-$ & $1.653$ & $1.560$ & $0.944$ \\
$0p_{1/2}^{-1}$ & $0d_{5/2}$ & $3^-$ & $1.951$ & $1.857$ & $0.952$ \\
$0p_{1/2}^{-1}$ & $1s_{1/2}$ & $0^-$ & $0.902$ & $0.641$ & $0.710$ \\
$0p_{1/2}^{-1}$ & $1s_{1/2}$ & $1^-$ & $1.179$ & $0.834$ & $0.707$ \\
\hline\hline
\end{tabular}
\end{center}
\end{table}
%\clearpage
%\begin{figure}
%%\epsfxsize=7cm
%\epsfysize=9.0cm
%\centerline{\epsffile{TES16_1.eps}}
%\vspace{2mm}
%\caption{Experimental energy spectra of
%the $^{16}$N--$^{16}$F mirror nuclei.}
%\label{fig:16N-F}
%\end{figure}
\end{document}